\documentclass{article}
\usepackage{fullpage}
\usepackage{spconf,amsmath}
\usepackage{amsthm}
\usepackage{amsfonts}
\usepackage{amssymb}
\usepackage{mathrsfs}
\usepackage{mathtools}
\usepackage[english]{babel}
\usepackage{float}
\usepackage[toc,page]{appendix}
\usepackage[pdftex]{graphicx}
\usepackage{epstopdf}
\usepackage{amsmath}
\usepackage{color}
\usepackage{multirow}
\usepackage{graphicx}
\usepackage{capt-of}

\usepackage{IEEEtrantools}
\usepackage{mdwlist}
\usepackage{microtype}
\usepackage[caption=false,font=footnotesize]{subfig}
\usepackage{cite}

\numberwithin{equation}{section}

\ninept

\newdimen\origiwspc%
\newdimen\origiwstr%
\origiwspc=\fontdimen2\font
\origiwstr=\fontdimen3\font
  
\title{Online Dictionary Learning Aided Target Recognition In Cognitive GPR}

\name{Fabio Giovanneschi$^{1}$, Kumar Vijay Mishra$^2$, Maria Antonia Gonzalez-Huici$^1$, Yonina C. Eldar$^2$,\thanks{The authors would like to thank the colleagues at Leibniz Institute for Applied Geophysics (LIAG), Hannover, Germany for their support during the measurement campaign. K.V.M. acknowledges partial support via Andrew and Erna Finci Viterbi Fellowship and Lady Davis Fellowship.}
}
\secondlinename{and Joachim H. G. Ender$^{1,3}$}
\address{$^1$Fraunhofer Institute for High Frequency Physics and Radar Techniques, Germany\\ \fontdimen2\font=0.3ex{$^2$Andrew and Erna Viterbi Faculty of Electrical Engineering, Technion - Israel Institute of Technology, Israel}\fontdimen2\font=\origiwspc \\$^3$Centre for Sensor Systems (ZESS), University of Siegen, Germany
}

\begin{document}

\maketitle

\begin{abstract}
Sparse decomposition of ground penetration radar (GPR) signals facilitates the use of compressed sensing techniques for faster data acquisition and enhanced feature extraction for target classification. In this paper, we investigate use of an online dictionary learning (ODL) technique in the context of GPR to bring down the learning time as well as improve identification of abandoned anti-personnel landmines. Our experimental results using real data from an L-band GPR for PMN/PMA2, ERA and T72 mines show that ODL reduces learning time by 94\% and increases clutter detection by 10\% over the classical K-SVD algorithm. Moreover, our methods could be helpful in cognitive operation of the GPR where the system adapts the range sampling based on the learned dictionary.
\end{abstract}


\vspace{-8pt}
\section{Introduction}
\label{sec:intro}
\vspace{-3pt}
A ground penetrating radar (GPR, hereafter) is used for probing the underground by transmitting radio waves in the subsurface and recording the backscattered reflections. The interest in GPR is due to its ability to reveal buried objects non-invasively and detect non-metallic scatterers with increased sensitivity to dielectric contrast \cite{jol2008ground,persico2014introduction}. This sensing technique is, therefore, attractive for several applications such as geophysics, archaeology, forensics, and defense (see e.g. \cite{daniels2005ground,jol2008ground} for some surveys). In this work, our focus is detection of buried landmines. It is one of the most extensively investigated GPR applications due to its obvious security and humanitarian importance. 

Mine detection GPR usually operates in L-band ($1$-$2$ GHz) with ultra-wideband (UWB) transmit signals that allow resolving small targets ($5$-$10$ cm diameter) at shallow depths ($~15$-$30$ cm) \cite{yarovoy2002ultra, giovanneschi2013parametric}. In such situations, GPR target recognition suffers from signal distortion due to inhomogeneous soil clutter, surface roughness and antenna ringing. Moreover, the constituting material of many models of landmines is largely plastic and has a very weak response to radar signals due to its low dielectric contrast with respect to the soil \cite{daniels2005ground,bruschini2016survey}. A variety of signal processing algorithms have been proposed for detection of low metal-content landmines in realistic scenarios; approaches based on feature extraction and classification are found to be the most effective (see e.g. \cite{ratto2011exploiting,gonzalez2013combined,torrione2014histograms,giannakis2016model}), yet false-alarm rates remain very high. Further, a high-resolution GPR has long scan times thereby making the data acquisition by a portable instrument very cumbersome \cite{suksmono2010compressive}.

In order to reduce the scan time or number of measurements, an emerging trend in GPR research \cite{gurbuz2012compressive,krueger2014compressive} is to employ the recently proposed compressed sensing (CS) framework \cite{eldar2012compressed,eldar2015sampling}. In CS, a signal can be reconstructed using a reduced number of samples w.r.t. the the Nyquist rate requirements, provided the signal is sparse in some domain.However, unlike point scatterers, the mine echoes are spatially extended and the resulting GPR received signal is not sparse in conventional range-time and frequency domains \cite{shao2013sparse}. 
Therefore, our immediate goal is to find an efficient sparse representation (SR) which accurately represents the scattering behaviors related to soil type and targets. This has been shown to improve the classifier performance in discriminating mines from clutter \cite{shao2013sparse, giovanneschi2015preliminary}.

In SR, the signal-of-interest is transformed to a domain where the signal can be expressed as a linear combination of only a few columns or \textit{atoms} of the \textit{dictionary} matrix \cite{elad2006image}. When it is inefficient to pre-define the dictionary to contain arbitrary basis (e.g. Fourier or wavelets), the usual resort is to \textit{learn} the dictionary from previous measurements. Dictionary learning (DL) techniques aim to create adapted dictionaries which provide the sparsest reconstruction for given training-sets, i.e., a representation with a minimum number of constituting atoms. Classical DL algorithms such as Method of Optimal Directions (MOD) \cite{engan1999method} and K-SVD \cite{elad2006image} operate in batches - dealing with the entire training set in each iteration. Although extremely successful, these methods are computationally demanding and not scalable to high-dimension training sets. An efficient alternative is the online dictionary learning (ODL) algorithm \cite{mairal2009online} that converges fast, processes small sets, and can infer the dictionary from large or time-varying training sets \cite{naderahmadian2016correlation}. 

Improved DL methods can aid in better target identification and subsequent reduction in GPR measurements through CS-based design. To this end, our work focuses on hitherto unexamined application of DL towards GPR-based landmine classification. Only one previous study has employed DL (K-SVD) using GPR signals \cite{shao2013sparse}, for identifying bedrock features. We propose employing ODL and then use the coefficients of the resulting sparse vectors as input to a Support Vector Machine (SVM) classifier to distinguish mines from clutter. Our comparison of ODL and K-SVD using real data from L-band GPR shows that ODL enjoys distinct advantage in speed and low false-alarm rates. Fast ODL computations pave the way towards cognitive GPR operation, wherein the system uses previous measurements to optimize the processing performance and is capable of sequential sampling adaptation based on the learned dictionary \cite{guerci2010cognitive,mishra2016cognitive,mishra2017performance}.

In the next section, we describe the GPR system and the data sets. In Section III, we introduce our technique for GPR target identification, with particular focus on ODL. Section IV presents classification and reconstruction results using real radar data. We provide concluding remarks in Section V.
\vspace{-6pt}
\section{System and field tests}
\label{sec:sys_meth}
\vspace{-3pt}
We used a commercial GPR system and carried out the field campaign at Leibniz Institute for Applied Geophysics (LIAG), Hannover (Germany) \cite{gonzalez2013combined}. We now provide details for the system and data.
\vspace{-8pt}
\subsection{L-band GPR}
\label{subsec:gpr}
The GPR system (see Fig. \ref{fig:GPR} inset) is an impulse radar with central frequency of 2 GHz.The frequency of the pulse repetition (PRF) and the sampling of the receiver ADC is of $1$ MHz. Table \ref{tbl:techparams} lists the salient technical parameters of the system. The scan rate of the system is $\sim1$ m/s with sampling resolution of 1 cm towards the perpendicular broadside (or X direction) and $4$ cm towards the cross-beam (Y direction). The radar uses a $8$ cm$\times8$ cm dual bow-tie dipole antenna for both transmit (Tx) and receive (Rx) sealed in a metallic shielding and an internal absorber. 
The raw data consists of samples of complex envelope of the received signal.
\begin{table}[t]
\begin{minipage}[b]{0.5\linewidth}
	\centering
	\includegraphics[scale=0.2]{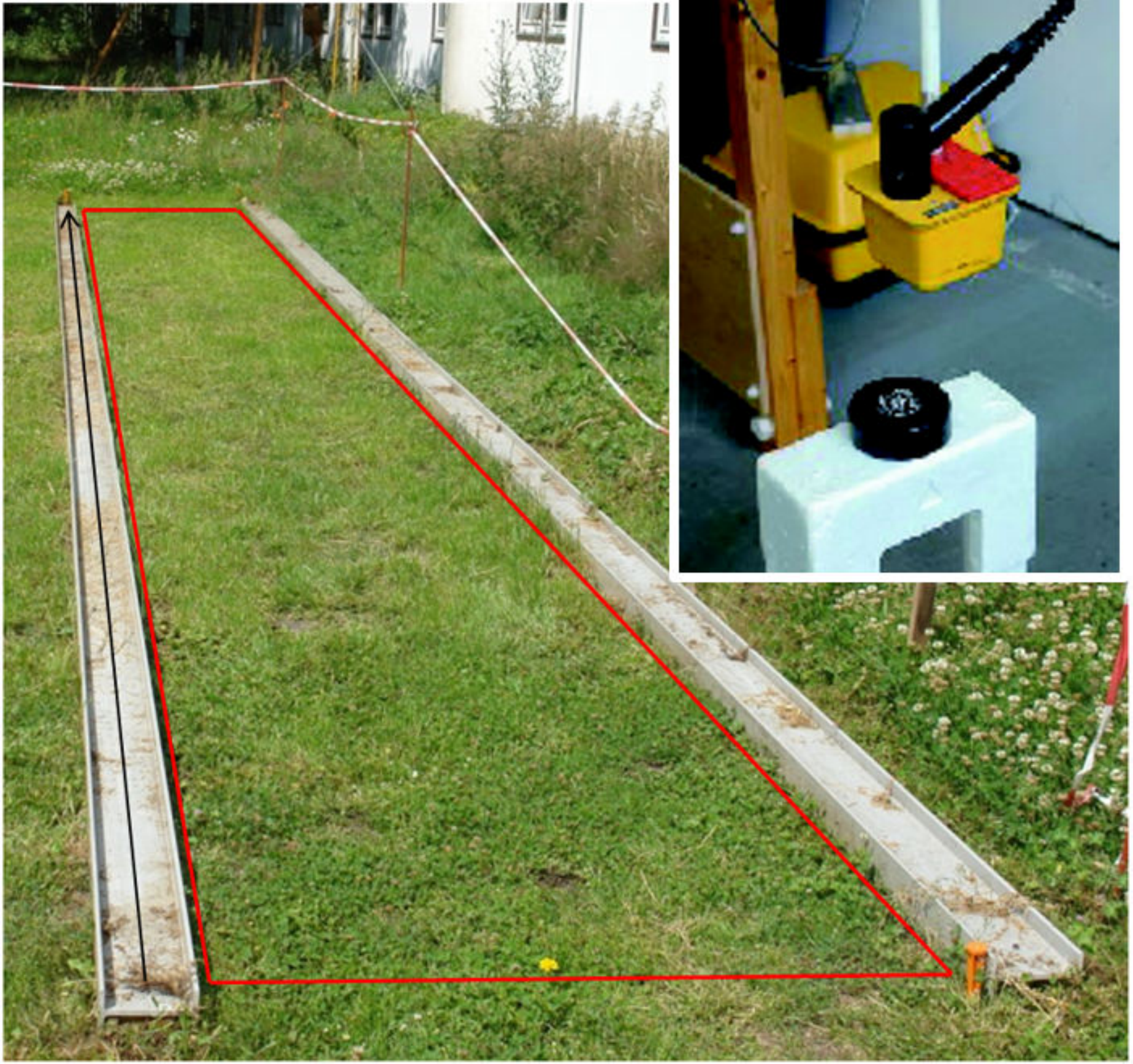}
  	\captionof{figure}{Test site (red) with buried mines. Inset shows GPR system.}
	\label{fig:GPR}
\end{minipage}\hfill
\begin{minipage}[b]{0.47\linewidth}
\centering
	\begin{tabular}{ l | l }
		\hline
         \noalign{\vskip 1pt}    
         	Parameter & Value\\[1pt]
		\hline
		\hline
        \noalign{\vskip 1pt}    
		   	Operating frequency & 2 GHz\\[1pt]
		   	PRF & 1 MHz\\[1pt]
		   	Pulse length & 0.5 ns\\[1pt]            
            Sampling time & 25 ps\\[1pt]
            Spatial sampling & 1 cm\\[1pt]
            Cross resolution & 4 cm\\[1pt]
            Antenna height & 5-9 cm\\[1pt]
            Samples/A-scan & 512\\
		\hline
		\hline
	\end{tabular}
    \caption{GPR parameters}
	\label{tbl:techparams}
\end{minipage}
\vspace{-14pt}
\end{table}
\vspace{-8pt}
\subsection{Mines data}
\label{subsec:minesdata}
The testbed was a grassy, moderately rough surface containing landmine simulants of different sizes that in the order of decreasing size are PMN/PMA2, ERA and T72, all buried at a depth of $5$-$10$ cm \cite{gonzalez2013accurate}. During the field tests, the GPR scanned different $1$ m$\times1$ m sections of the test-bed. The soil texture was sandy and highly inhomegeneous (due to the presence of material such as organic matter and stones), thereby leading to a high variability in the electrical parameters. We measured the dielectric constant at three different locations of the testbed with a Time Domain Reflectometer (TDR) to obtain an estimate of its mean value and variability. The average value oscillated between 4.6 and 10.1 with $15\%$ standard deviation and correlation length \cite{gonzalez2013combined} of $20$ cm. Such big variations in soil compositions pose difficulties in mine detection with existing methods.
\vspace{-6pt}
\section{GPR Target Identification Method}
\label{sec:methodology}
\vspace{-3pt}
We now describe our method for dictionary learning and classification for GPR target identification. The literature for DL and SVM is extensive, and hence we only summarize these methods here. In the following, we use boldface lowercase and uppercase letters for vectors and matrices, respectively.
\vspace{-8pt}
\subsection{Dictionary learning}
A dictionary learning algorithm finds an over-complete dictionary $\mathbf{D} \in \mathbb{R}^{m\times n}$, $m<n$ that can sparsely represent measurements $\mathbf{y}\in \mathbb{R}^m$. For the training data $\mathbf{Y} = [\mathbf{y}_1, \cdots, \mathbf{y}_L]$, we call $\mathbf{X} = [\mathbf{x}_1, \cdots, \mathbf{x}_L] \in \mathbb{R}^{n\times L}$ a sparse representation of $\mathbf{Y}$ over $\mathbf{D} = [\mathbf{d}_1, \cdots, \mathbf{d}_n]$, if $\mathbf{Y} \simeq \mathbf{D}\mathbf{X}$. Here, $\mathbf{d}_j$, $j=1,\cdots,n$ is called an atom of the dictionary. Each of the vectors $\mathbf{x}_i$ is a sparse representation of $\mathbf{y}_i$ with only $K$ nonzero entries. A tractable formulation of this problem is the following non-convex optimization\par\noindent\small
\begin{flalign}
\label{eq:csrecoverl1_1}
	& \underset{\mathbf{D},\mathbf{X}}{\text{minimize}}\phantom{1}\left\Vert \mathbf{Y}-\mathbf{D}\mathbf{X}\right\Vert _{F}\nonumber\\
	& \text{subject to}\phantom{1} \left\Vert\mathbf{x}_i\right\Vert_0 \le K,\:\forall 1\le i \le L,
\end{flalign}\normalsize
where $||\cdot||_F$ denotes Frobenius norm. Since both $\mathbf{D}$ and $\mathbf{X}$ are unknown, commonly this is turned into a two-step convex problem that alternately minimizes $\mathbf{X}$ (\textit{sparse coding step}) \cite{aharon2006k} and $\mathbf{D}$ (\textit{dictionary update step}). The popular K-SVD algorithm sequentially updates all the atoms 
for each alternation between the aforementioned two steps. This is a \textit{batch} algorithm that uses the entire training data for updating the dictionary at each alternation. The ODL also updates the entire dictionary sequentially, but uses one element of training data at a time for the gradient descent-based dictionary update step. For the sparse coding step, K-SVD employs Orthogonal Matching Pursuit (OMP) with the formulation: \par\noindent\small
\begin{flalign}
\label{eq:omp}
& \underset{\mathbf{x}_i}{\text{minimize}}\phantom{1}\left\Vert\mathbf{x}_i\right\Vert_0\nonumber\\
	& \text{subject to}\phantom{1} \left\Vert \mathbf{y}_i-\mathbf{D}\mathbf{x}_i\right\Vert^2_{2} \le \alpha,\:\forall 1\le i \le L,
\end{flalign}\normalsize
where $\alpha$ is the maximum residual error used as a stopping criterion. The ODL, on the other hand, uses Cholesky-based implementation of the LARS-LASSO algorithm \cite{osborne2000new}. The latter solves a $\ell_1$-regularized least-squares problem:\par\noindent\small
\begin{align}
\label{eq:lars}
\mathbf{x}_T \triangleq  \underset{\mathbf{x}\in \mathbb{R}^n}{\mathrm{min}} \frac{1}{2}||\mathbf{y}_T - \mathbf{D}_{T-1}\mathbf{x}||^2_2 + \lambda||x||_1,
\end{align}\normalsize
where the subscripts $T-1$ and $T$ denote last and current iterations.
\vspace{-6pt}
\subsection{Signal classification}
We use SVM to classify the sparsely represented GPR range profiles. Given a pre-defined collection of labeled observations, namely a ``classification set'', SVM searches for a functional $f:\mathbf{R}^n \rightarrow \mathbf{R}$ that maps any given observation $\mathbf{x}_i$ to a class $c\in \mathbb{R}$. 
In our work, the classification set is a labeled collection of  GPR range profiles of targets/clutter which have been sparsely decomposed using the learned dictionary $\mathbf{D}$. We refer the reader to \cite{chang2011libsvm} for details of SVM. Briefly, SVM transforms the data into a high dimensional feature space where it is easier to separate between different classes. The kernel function that we use to compute the high dimensional operations in the feature space, is the Radial Basis Function (RBF).
 To optimally select the SVM input parameters, we arranged the original classification set into training and validation vectors in $\nu$ different ways ($\nu$-fold cross-validation with $\nu$=10) to arrive at a certain mean cross-classification accuracy of the validation vectors.

\vspace{-6pt}
\section{Experimental Results and Analysis}
\label{sec:results}
\vspace{-3pt}
We divided the entire LIAG data into three different sets: the training set to learn the dictionary, the classification set for the SVM classifier and the test set. We devised a statistical approach to select the best parameters for DL with GPR mines data, and then use the resultant dictionary for the classification procedure.
\vspace{-6pt}
\subsection{Dictionary comparison}
\label{subsec:compexp}
\begin{figure}[!t]
\centering
  \includegraphics[scale=0.25]{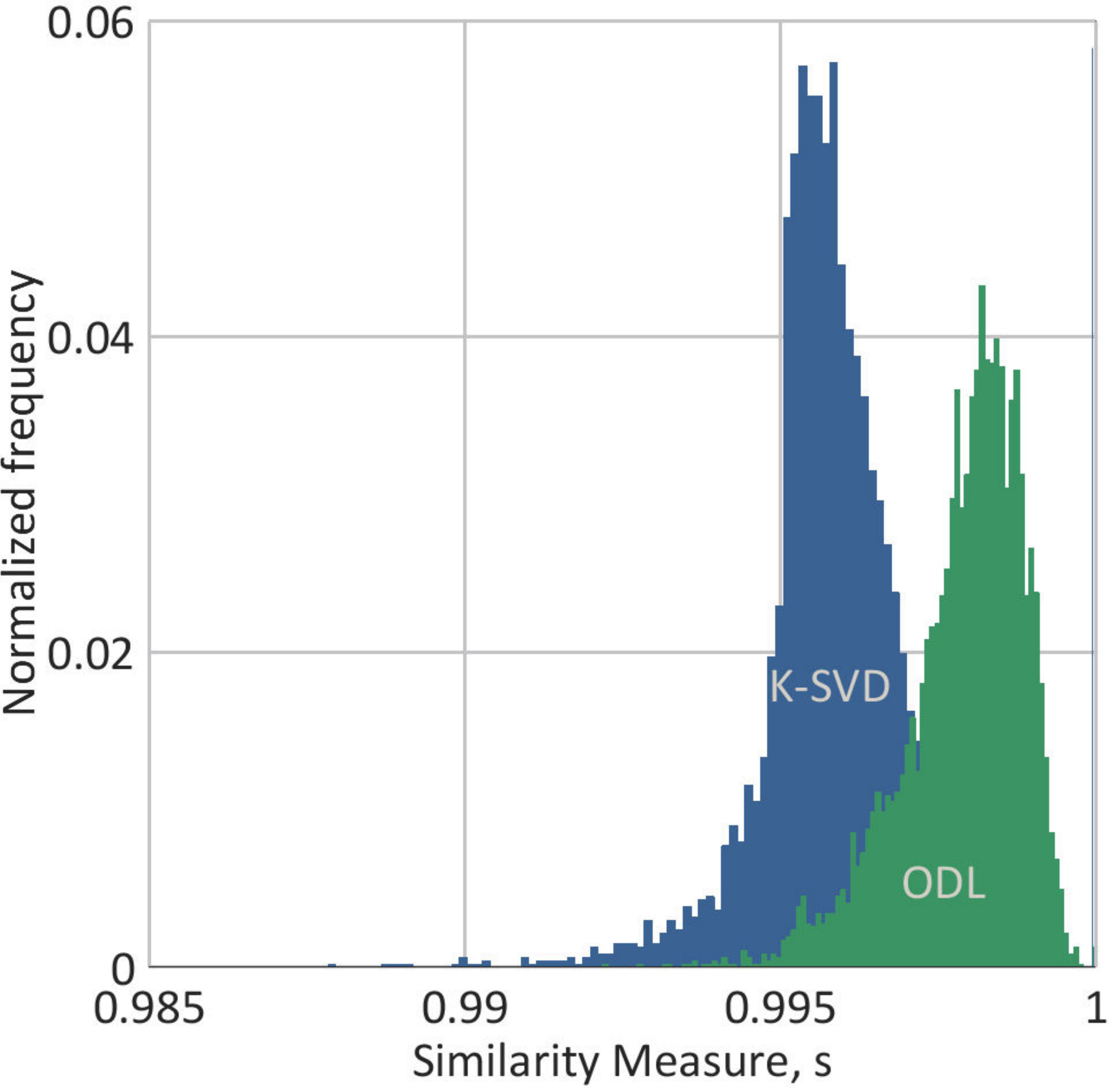}
  \caption{\scriptsize{Normalized histograms of similarity measure. Here, $T = 40$, $K=512$, $\lambda = 0.1$, and $\alpha = 0.1$.}\vspace{-9pt}}
\label{fig:odlKsvdDist}
\end{figure}
The training set $\mathbf{Y}$ consisted of more than two thousand range profiles that were randomly selected from a collection of GPR responses in proximity of the buried targets; these areas belonged to different surveys of the test site. In order to compare the dictionaries obtained from ODL and K-SVD, we use a \textit{similarity measure} that quantifies the closeness of the reconstructed vectors $\mathbf{\hat{y}}_i$ obtained using the sparse coefficients of the learned dictionary $\mathbf{D}$ with the original training set vectors $\mathbf{y}_i$. Let the cross-correlation between the two vectors be $\mathbf{r}_{\mathbf{y}_i,\mathbf{\hat{y}}_i}(m) = \sum\limits_{n=-\infty}^{+\infty}\mathbf{y}_i(n)\mathbf{\hat{y}}_i(n+m)$. The normalized cross-correlation is\par\noindent\small
\begin{align}
\overline{\mathbf{r}_{\mathbf{y}_i,\mathbf{\hat{y}}_i}}(m) = \frac{\mathbf{r}_{\mathbf{y}_i,\mathbf{\hat{y}}_i}(m)}{\sqrt{\mathbf{r}_{\mathbf{y}_i,\mathbf{y}_i}(0)\mathbf{r}_{\mathbf{\hat{y}}_i,\mathbf{\hat{y}}_i}(0)}}
\end{align}\normalsize
We define the similarity measure $s_i$ to be the maximum of the absolute value of the normalized cross correlation between $\mathbf{y}_i$ and $\mathbf{\hat{y}}_i$: $s_i = \textrm{max}|\overline{\mathbf{r}_{\mathbf{y}_i,\mathbf{\hat{y}}_i}(m)}|$. The set of similarity measure values $\{s_i\}_{i=1}^{N}$ form an empirical probability density function (epdf), $p_{s_{DL}}$ where the subscript DL represents the method used (K-SVD or ODL). As an example, Fig. \ref{fig:odlKsvdDist} shows the epdfs for the dictionaries learned for the GPR mines data. We note that $p_{s_{ODL}}$ is more skewed towards unity than $p_{s_{KSVD}}$, thereby demonstrating similarity of the learned ODL dictionary with the original training set. To arrive at optimum parameter values affecting these epdfs, we now use statistical metrics.
\vspace{-8pt}
\subsection{Parameter analysis}
\label{subsec:paranal}
\begin{figure}[!t]
\centering
\subfloat[Coefficient of variation]{%
  \includegraphics[width=3.9cm]{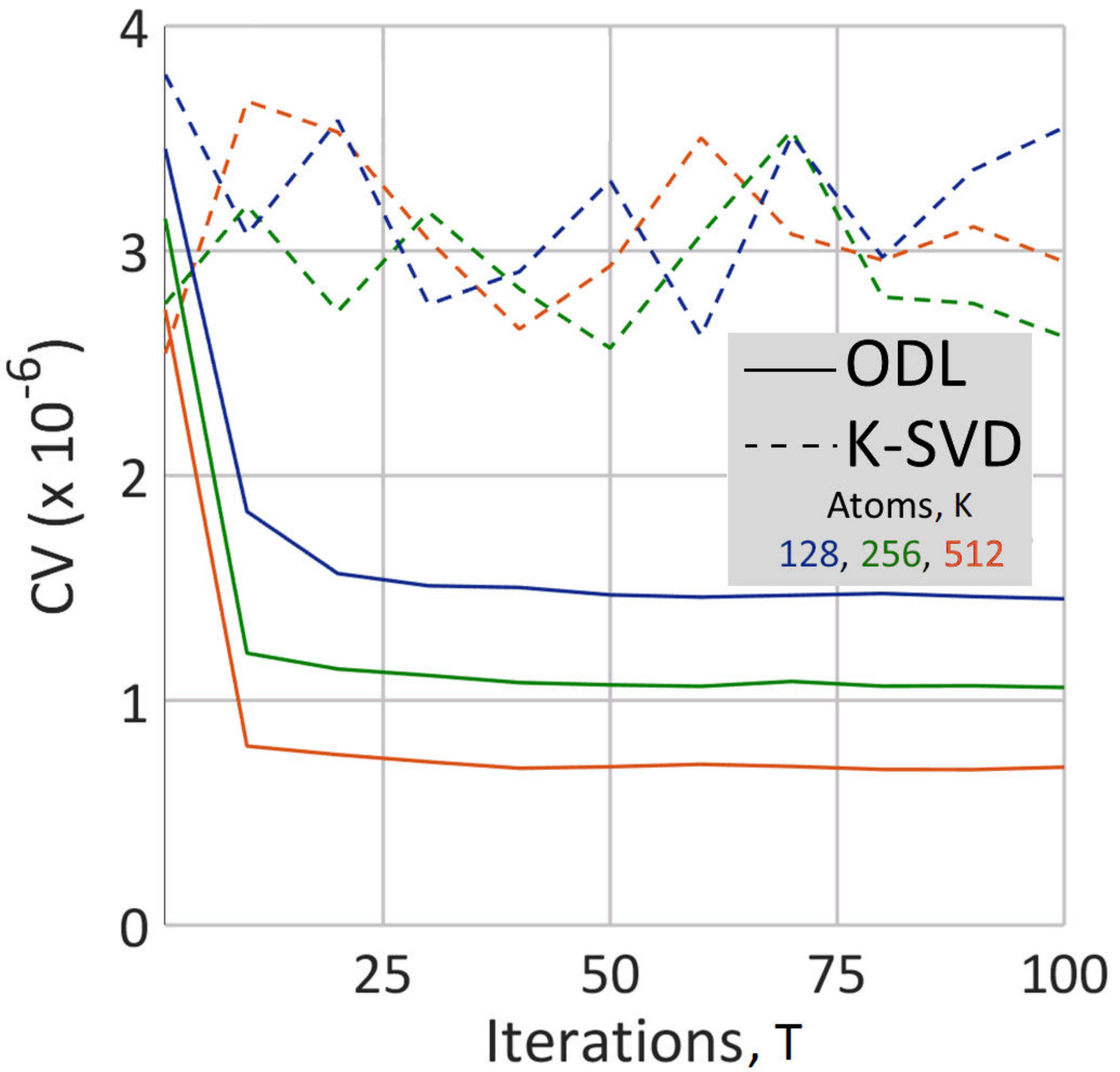}%
  \label{fig:cov}%
}\qquad
\subfloat[Kolmogorov-Smirnov distance]{%
  \includegraphics[width=4.0cm]{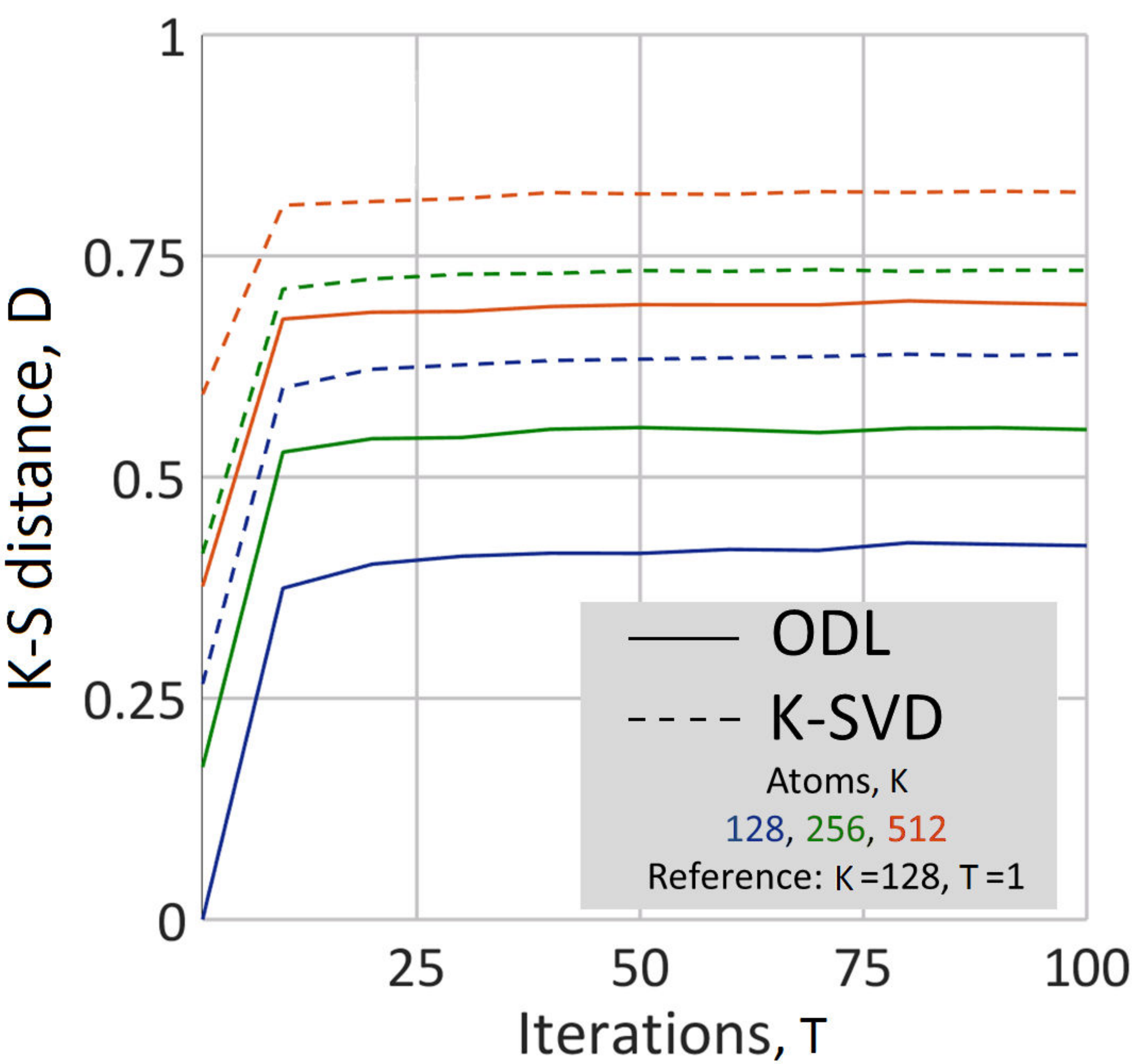}%
  \label{fig:ksd}%
}
\caption{\scriptsize{Statistics of similarity measure distributions for $\lambda=0.1$ and $\alpha=0.1$ and varying number of iterations and trained atoms.}\vspace{-14pt}}
\label{fig:cvks}
\end{figure}
The parameters which predominantly affect the results of ODL and K-SVD algorithms are the number of iterations ($T$), the number of trained atoms $K$, the regularization parameter $\lambda$ in the SD step, and the error parameter $\alpha$ to sparsely decompose the training set via OMP. The epdf $p_s$ is then a function of these four parameters. For a given DL method, our goal is to compare the epdfs of similarity measure by varying these parameters, and arrive at the thresholds of parameter values after which the changes in $p_{s_{DL}}$ are only incremental. 

We are looking for set of values $\{T, K, \lambda, \alpha\}$ for which the $p_{s_{DL}}$ is skewed towards unity and has small variance. The individual comparisons of mean ($\mu$) and standard deviation ($\sigma$), as used in previous GPR DL studies \cite{shao2013sparse}, are not sufficient to quantify the observed dispersion in the epdfs obtained by varying any of the parameter values. We, therefore, simultaneously compare both statistics by using the coefficient of variation, $CV = \sigma/\mu$; in our analysis, it represents the extent of variability in relation to the mean of the similarity values. Figure \ref{fig:cov} shows the variation of CV for ODL and K-SVD for different values of trained atoms and fixed values of $\lambda=0.1$ and $\alpha=0.1$. The ODL shows that after $40$ iterations, the variation in the epdf is negligible and the impact of increasing the number of trained atoms grows. We note that K-SVD doesn't show any such trend.

Our second metric to compare the distributions of the similarity measure obtained by successive changes in parameter values is the Kolmogorov-Smirnov (K-S) distance \cite{chakravarti1967handbook}, which is the maximum distance between two given empirical cumulative distribution functions (ecdf). Larger values of K-S distance indicate that samples are drawn from different underlying distributions. Suppose $P_{s_1}$ and $P_{s_2}$ are the ecdfs of the same length corresponding to epdfs $p_{s_1}$ and $p_{s_2}$, respectively. Then K-S distance $D(P_{s_1}, P_{s_2})$ is\par\noindent\small
\begin{align}
D(P_{s_1}, P_{s_2}) = \sup_{1 \le i \le N}|P_{s_1}(i) - P_{s_2}(i)|,
\end{align}\normalsize
where $\sup$ denotes the supremum over all distances. Figure \ref{fig:ksd} shows the variation of K-S distance for the identical values of parameters as used for CV evaluation in Fig. \ref{fig:cov}. We kept the ecdf corresponding to $K=128$ atoms and $T=1$ iteration as a reference. We then computed the K-S distance of ecdfs obtained with other parameter values with respect to this reference. As the parameter values move away from this reference, the K-S distance increases. However, for both DL methods, we notice the following trend: as the number of iterations increase, the difference in K-S distance between successive iterations is negligible and is influenced mostly by the number of trained atoms. The K-S distance quantifies the difference between ODL and K-SVD distributions rather than stating which one is better. Combining this information with Fig. \ref{fig:cov}, it is evident that ODL has lower CV and is also more robust to parameter changes than K-SVD.
\vspace{-8pt}
\subsection{Classification and computational efficiency}
\label{subsec:compeff}
After examining the thresholds at which the CV and K-S distance stabilize, we selected the following values for dictionary learning: $T = 40$, $K=512$, $\lambda = 0.1$, and $\alpha = 0.1$. For this set of values, Fig. \ref{fig:odlKsvdDist} shows the epdfs of similarity measures for ODL and K-SVD. In a landmine clearance campaign, these parameters will be learned offline and will remain fixed during the classification procedure. The test set contains 2 different surveys for each type of mine class for a total of five thousand range profiles. Once the test-set is sparsely decomposed using the learned dictionary, we estimated a mean sparsity to be $\approx 4$ for ODL and $\approx 6$ for K-SVD. 
In order to find the best functional for the SVM classifier, we conducted a cross-validation on the sparsely decomposed classification set using a set of different RBF kernel parameters. After obtaining the functional, we used it to classify the sparsely decomposed test set. 

\begin{figure}[!t]
\centering
  \includegraphics[scale=0.25]{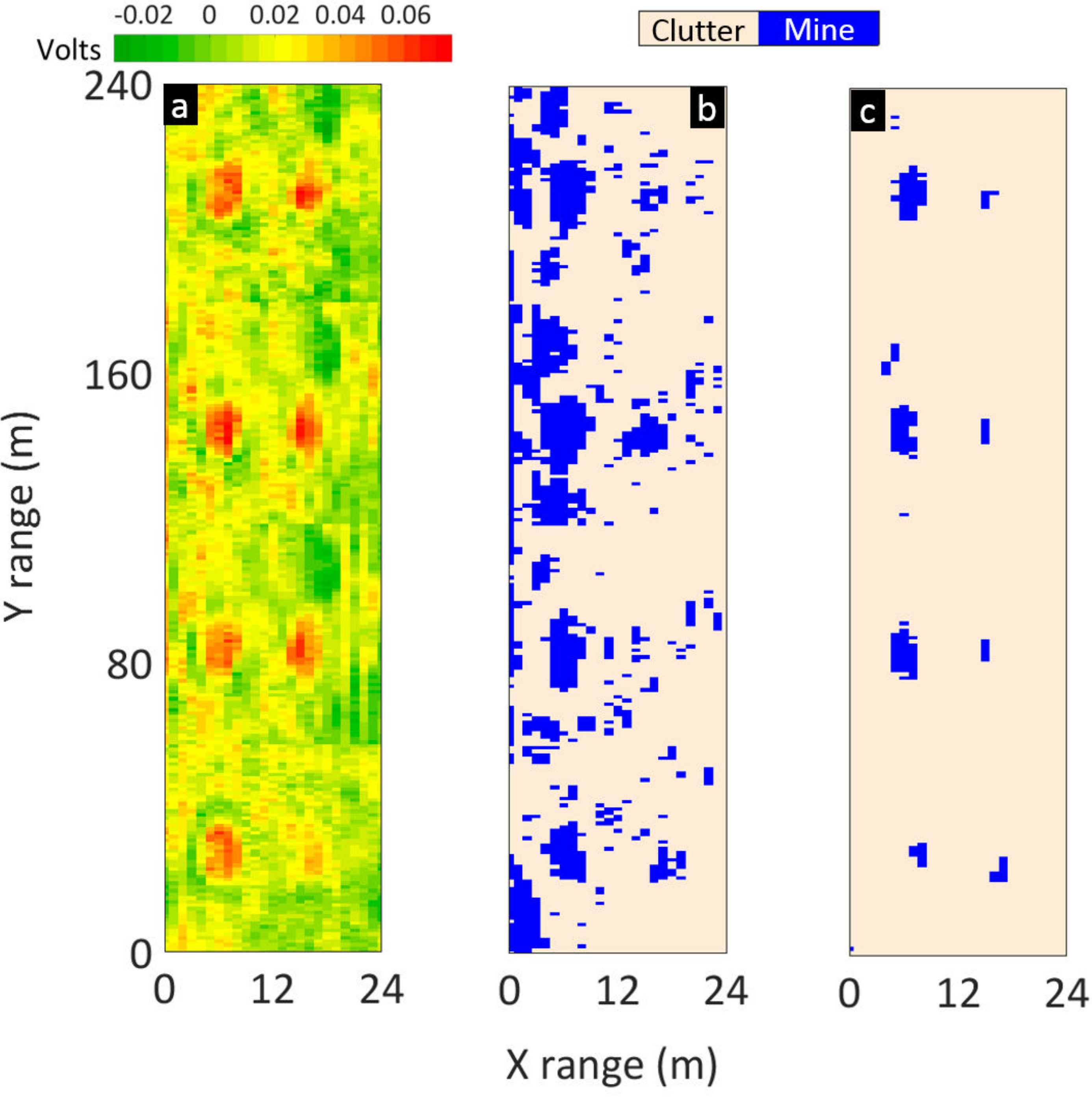}
  \caption{(a) Raw-data for PMA/PMN (warmer values indicate presence of mines). Classification map obtained with (b) K-SVD and (c) ODL.\vspace{-10pt}}
\label{fig:classResult}
\end{figure}

Figure \ref{fig:classResult} shows the raw data for PMA/PMN mines at $15$ cm depth and the classification maps obtained with a learned dictionary using ODL and K-SVD methods. It is clear that target and clutter recognition are drastically improved using a dictionary learned with ODL when compared with K-SVD.
Table \ref{tbl:perf} compares the performance of classification using the two DL methods. Using accurate ground truth information, we defined \textit{target halos} as the boundaries of the buried landmines. Let the number of pixels and the declared mine pixels inside the target halo be $N_t$ and $N_m$, respectively. Similarly, we denote the number of true and declared clutter pixels outside the target halo by $N_c$ and $N_d$, respectively. Then, the probabilities of correct classification ($P_{CC}$) for mines and clutter are
\begin{align}
P_{CC_{\text{mines}}} = \frac{N_m}{N_t},\:\:\textrm{and}\:\:P_{CC_{\text{clutter}}} = \frac{N_d}{N_c}.
\end{align}
The $P_{CC}$ being the output of a classifier should not be mistaken as the radar's probability of detection $P_d$ which is the result of a detector. A detector would declare the presence of a mine when only a few pixels inside the halo have been declared as mine. Since $P_{CC}$ takes into account the entire target halo, it provides a fair and accurate evaluation of of the classification result.
The execution time for sparse decomposition and classification steps were identical ($0.26$ s and $0.14$ s, respectively) for both ODL and K-SVD methods. However, for DL update, ODL took only $0.46$ s - more than sixteen times faster than K-SVD ($8.09$ s).
\begin{table}[t]
\centering
	\begin{tabular}{ l | l | l | l | l}
		\hline
         \noalign{\vskip 1pt}
           	\multirow{ 2}{*}{} & \multicolumn{4}{c}{$P_{CC}$}\\
            \cline{2-5}
        \noalign{\vskip 1pt}
            & Clutter & PMN/PMA2 & ERA & T72\\[1pt]
		\hline
        \hline
        \noalign{\vskip 1pt}
            K-SVD & 0.824 & 0.810 & 0.666 & 0.750 \\[1pt]
            ODL & 0.923 & 0.950 & 0.833 & 0.750 \\
		\hline
		\hline
	\end{tabular}
    \caption{\small{Performance of ODL and K-SVD for mine-detection GPR. ODL and KSVD results are identical for the smallest mine T72 because only a few pixels are considered for T72 classification.}\vspace{-14pt}}
	\label{tbl:perf}
\end{table}    
\vspace{-6pt}
\section{Summary}
\label{sec:summary}
In this work, we proposed ODL for sparse decomposition of GPR-based mine data. Our results indicate near-real-time execution times using ODL, high clutter rejection and improved classifier performance. These characteristics open interesting opportunities for future cognitive operation of GPR. For example, in a realistic landmine-clearance campaign, an operator could gather the training measurements over a safe area next to the contaminated site, hypothetically placing some buried landmine simulants over it in order to have a faithful representation of the soil/targets interaction beneath the surface. In other words, our work allows the operator to ``calibrate'' the acquisition by providing a good training set to learn the dictionary. 

\newcommand{\BIBdecl}{\setlength{\itemsep}{0.001 em}}
\bibliographystyle{IEEEtran}
\scriptsize{\bibliography{refs}}

\end{document}